\documentclass[sigconf,dvipsnames,authorversion]{acmart}

\usepackage{xcolor}
\usepackage{algorithmic}
\usepackage{graphicx}
\usepackage{textcomp}
\usepackage{xcolor}
\usepackage{booktabs}
\usepackage{siunitx}
\usepackage{caption}
\usepackage{subcaption}

\setcopyright{rightsretained}

\acmDOI{10.1145/3605098.3635938}

\acmISBN{979-8-4007-0243-3/24/04}

\acmConference[SAC '24]{The 39th ACM/SIGAPP Symposium on Applied Computing}{April 8--12, 2024}{Avila, Spain}
\acmYear{2024}
\copyrightyear{2024}

\acmPrice{15.00}

\begin{document}
\title{Classification of Home Network Problems with Transformers}

\author{Jeremias Dötterl}
\orcid{0009-0007-7801-4639}
\affiliation{%
  \institution{Adtran}
  \streetaddress{Hermann-Dorner-Allee 91}
  \city{Berlin} 
  \country{Germany}
}
\email{jeremias.doetterl@adtran.com}

\author{Zahra Hemmati Fard}
\orcid{0009-0004-3858-2015}
\affiliation{%
  \institution{Adtran}
  \streetaddress{Hermann-Dorner-Allee 91}
  \city{Berlin} 
  \country{Germany}
}
\email{zahra.hematifard@adtran.com}

\begin{abstract}
We propose a classifier that can identify ten common home network problems based on the raw textual output of networking tools such as ping, dig, and ip.
Our deep learning model uses an encoder-only transformer architecture with a particular pre-tokenizer that we propose for splitting the tool output into token sequences.
The use of transformers distinguishes our approach from related work on network problem classification, which still primarily relies on non-deep-learning methods.
Our model achieves high accuracy in our experiments, demonstrating the high potential of transformer-based problem classification for the home network.
\end{abstract}

\ccsdesc[500]{Networks~Home networks}
\ccsdesc[300]{Networks~Network reliability}
\ccsdesc[100]{Networks~Network monitoring}
\ccsdesc[500]{Computing methodologies~Neural networks}

\keywords{Network troubleshooting, fault classification, machine learning, deep learning, transformers}

\maketitle

\section{Introduction}
Network problems are a common and frustrating experience for subscribers, who expect their network to function reliably.
Often, problems are caused in the home network, e.g., by router misconfigurations, poor WiFi signal, or DNS-related issues.
The root cause of such problems is difficult to identify for subscribers, who, therefore, regularly require technical support from their service providers.
For service providers, offering such tech support is costly, and automation software that can reliably classify and repair common home network problems could significantly reduce these costs.

Network problems can be diagnosed using specialized tools, rule-based systems, or machine learning \cite{tong2018}.
Specialized tools include Linux commandline utilities like ping, dig, ip, and ethtool, which are mature, widely available, and highly valued by networking experts.
Rule-based systems encode expert knowledge with \emph{if-then} rules to (partially) automate problem diagnosis.
But as accurate rules are difficult to develop and maintain, machine learning solutions have become popular, often using decision trees, random forests, or support vector machines.

A promising idea is to combine tool-based diagnosis with ma\-chine-learning-based diagnosis by feeding the output of networking tools into machine learning models.
However, conventional machine learning algorithms cannot work on text directly, which makes them difficult to apply to the textual tool output.
Ingesting the tool output into these algorithms requires extensive parsing and feature engineering, which is cumbersome and error-prone and additional programming effort is necessary every time a new tool should be added.

In recent years, deep learning \cite{lecun2015} has enabled big advances in text processing;
sequential deep learning models are designed to work on text directly and can be leveraged for text classification \cite{sutskever2014}.
In particular, the transformer architecture \cite{vaswani2017} has marked an important milestone as it has demonstrated outstanding performance in many application scenarios and can be applied to sequential input of varying length without the complexity of Long Short-Term Memory networks (LSTMs).
This makes transformers an interesting candidate for our use case.

In this paper, we follow a simple idea:
We train a transformer model that can classify ten common home network problems based on the raw textual output of networking tools such as ping, dig, and ip.
Our model pipeline consumes concatenated strings and outputs probability distributions over the potential problem classes.
Such an end-to-end sequential deep learning model avoids the complexity and costs of parsing and manual feature engineering.

To boost model performance, we propose a pre-tokenizer named \emph{Greedy-k-digits} for splitting tool output into token sequences, which is more suitable for our use case than the pre-tokenizers prevalent in natural language processing.
The pre-tokenizer splits numbers after at most $k$ digits and helps the model achieve high accuracy in our experiments, demonstrating the large potential of transformer-based network problem classification.

The rest of this paper is structured as follows.
In section~\ref{sec:related-work}, we discuss related work with a particular focus on machine learning for network problem classification.
Section~\ref{sec:common-problems} introduces common home network problems that we want to classify in this paper.
Section~\ref{sec:approach} presents transformer-based home network problem classification and the \emph{Greedy-k-digits} pre-tokenizer, which are the main contributions of our work.
In section~\ref{sec:simulation}, we introduce the home network simulation that we developed to generate a suitable dataset.
Section~\ref{sec:evaluation} presents our evaluation and experimental results. 
Finally, section~\ref{sec:conclusion} summarizes our conclusions.

\section{Related Work}
\label{sec:related-work}

Automating the classification of network problems is crucial for resolving them quickly and with minimal human effort.
It has been argued that such a classification is particularly valuable in the home network, because it is maintained by the subscriber who often does not possess the knowledge required to understand, analyze, and fix problems with the network \cite{dong2011,dalal2014}.
An autonomous classifier could not only assist the subscriber during troubleshooting, but also play an important role in future self-healing home networks \cite{dalal2014}.

Troubleshooting can be automated with rules or machine learning \cite{tong2018}.
An interesting rule-based technique is presented in \cite{dong2011}, which generates explanations for home network failures by reasoning on arguments and counterarguments.
Unfortunately, this approach requires high-quality rules, which must be contributed by networking experts.
This ''knowledge bottleneck'' has shifted research attention from rule-based to machine-learning-based solutions, which can derive (implicit) rules from labeled datasets.

Many prior works on network problem classification rely on non-deep-learning machine learning algorithms like decision trees, random forests, and support vector machines.
For example, decision trees are used by Moulay et al.~\cite{moulay2020} to classify anomalies in TCP KPIs and by Chen et al.~\cite{chen2004} to classify faults in large website logs at Ebay.
Support vector machines, random forests, and neural networks are used by Srinivasan et al.~\cite{srinivasan2019} to detect and localize link faults.
Madi et al.~\cite{madi2020} compare the accuracy of different classification algorithms (support vector machines, k-nearest neighbor, naive bayes, random forests, etc.) for network fault management using syslog data.
Syrigos et al.~\cite{syrigos2019} compare the performance of support vector machines, decision trees, random forests, and k-nearest neighbor for WiFi pathology classification.
One disadvantage of these algorithms is that they require feature engineering while sequential deep learning models can be trained on raw text.

While deep learning plays an increasingly bigger role in network traffic monitoring \cite{abbasi2021} and log-based anomaly detection \cite{landauer2023}, there seems to be little prior work on deep learning for log-based network problem classification.
The most relevant prior work appears to be by Ramachandran et al.~\cite{ramachandran2023}, who use deep learning to classify errors in system logs.
Their work uses a novel manual vectorization technique and LSTMs.
In contrast, in our work we train an end-to-end transformer model and evaluate how accurately home network problems can be classified based on the output of standard networking tools, which apparently has not been investigated.

One additional novelty of our approach is a new pre-tokenizer for splitting tool output into token sequences.
Most pre-tokenizers used in natural language processing split on whitespace characters and punctuation marks \cite{mielke2021}, which is appropriate for natural language but not ideal for tool output:
Splitting on whitespaces creates unnecessarily long token sequences, which require more complex models to work well.
Moreover, the most important information contained in tool outputs is often carried by numbers, hence our pre-tokenizer applies special treatment to numbers, which boosts the model accuracy in our experiments.

\section{Common Home Network Problems}
\label{sec:common-problems}

This section describes the four different categories of home network problems that we want to address in this paper.

\begin{itemize}
    \item \emph{WiFi-related problems}:
        Common WiFi problems are poor signal due to limited coverage or interference from other WiFi networks or non-WiFi sources \cite{syrigos2019} (like microwave ovens and Bluetooth) or due to poor placement of the WiFi access point.
        Access points can also be misconfigured \cite{alotaibi2016}, e.g. the transmit power may be too low for noisy environments.
    \item \emph{Routing-related problems}:
        Connectivity problems can occur when the routing is misconfigured \cite{tong2018}, e.g., a host has a wrong route or is missing a default gateway \cite{dong2011} and traffic is not routed properly.
    \item \emph{Link-related problems}:
        Connectivity problems can be caused by link failures \cite{tong2018}, disconnected links (the subscriber unplugged the cable) \cite{dong2011} or high packet loss, delay, and jitter. 
    \item \emph{DNS-related problems}:
        DNS-related problems occur when the DNS server address is wrong or not configured \cite{dong2011} and the DNS server cannot be reached.
\end{itemize}

We identified 10~specific problems that we want to classify with our machine learning model.
Together with the normal network state (in absence of problems), our classifier is supposed to distinguish 11~classes, which are listed in Table~\ref{tab:problem-classes}.
\begin{table}
\centering
\caption{Problem classes}
\label{tab:problem-classes}
\begin{tabular}{ll}
\toprule
Problem class & Short description  \\ \midrule
\texttt{\footnotesize CORRUPT\_DEFAULT\_ROUTE}  & Wrong route configured on host   \\
\texttt{\footnotesize DNS\_WRONG\_IP}  & Wrong DNS server IP address      \\
\texttt{\footnotesize HIGH\_DELAY}  & Link with high delay     \\
\texttt{\footnotesize HIGH\_JITTER}  & Link with high jitter  \\
\texttt{\footnotesize HIGH\_PACKET\_LOSS}  & Link with high packet loss   \\
\texttt{\footnotesize HOST\_INTERFACE\_DOWN}  & Host ethernet interface down    \\
\texttt{\footnotesize LOW\_AP\_TX\_POWER}  & Access point tx power too low    \\
\texttt{\footnotesize NO\_DEFAULT\_ROUTE}  & No default route configured on host    \\
\texttt{\footnotesize NORMAL\_STATE}  & No problem   \\
\texttt{\footnotesize ROUTER\_INTERFACE\_DOWN}  & Router ethernet interface down    \\
\texttt{\footnotesize STATION\_FAR\_AWAY}  & Station far away from access point     \\
\bottomrule
\end{tabular}
\end{table}

Most of the problem classes have similar symptoms (the Internet cannot be reached or the connection is slow) and are therefore difficult to distinguish for the non-expert subscriber, motivating the need for an automated problem classifier.

\section{Transformer-Based Network Problem Classification}
\label{sec:approach}

We are proposing a transformer-based network problem classifier, which is based on the observation that different network problems affect the output of networking tools in different ways.
We gather the tool output into log files and use transformers to learn an accurate mapping from log contents to problem classes.
While some of the simpler problem classes may be identified with simple bag-of-words models based on the presence and absence of certain keywords, we demonstrate in our experiments that transformers can identify difficult classes more accurately.

\begin{figure*}[t]
\centerline{\includegraphics[width=0.9\linewidth]{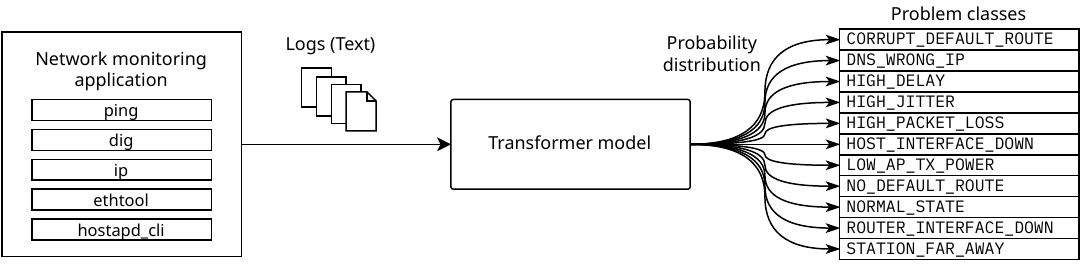}}
\caption{Transformer-based network problem classification}
\label{fig:transformer}
\end{figure*}

The high-level overview of our transformer-based network problem classification is shown in Fig.~\ref{fig:transformer}:
\begin{itemize}
    \item The \emph{network monitoring application}
        runs on the residential gateway (router and WiFi access point) of the home network and collects data about the network in regular intervals.
        This application can be a lightweight wrapper around the existing mature networking tools used by experts today:
        \begin{itemize}
            \item \emph{ping} is used to check whether a destination can be reached and to measure the roundtrip delay to the destination.
            \item \emph{dig} is used to query DNS servers.
            \item \emph{ip} is used to inspect the configured routes.
            \item \emph{ethtool} is used to check the ethernet link state.
            \item \emph{hostapd\_cli} is used to query the WiFi access point daemon hostapd for information like connected stations, signal strength, transmit power, etc.
        \end{itemize}
        The tool outputs are concatenated and this log entry is passed as a string to the transformer model.
    \item The \emph{transformer model} ingests the textual logs, performs inference, and outputs a probability distribution over the problem classes.
    These probabilities can then be used to inform the subscriber about the most likely problem or to attempt an automated repair.
\end{itemize}

The transformer model acts as a data transformation pipeline that converts strings to probability distributions.
We follow the original transformer architecture \cite{vaswani2017} closely, but we only use the encoder part of the architecture and instead of using a decoder, we pass the encoded output into a classification head.

\begin{figure}[ht]
\centerline{\includegraphics[width=0.8\linewidth]{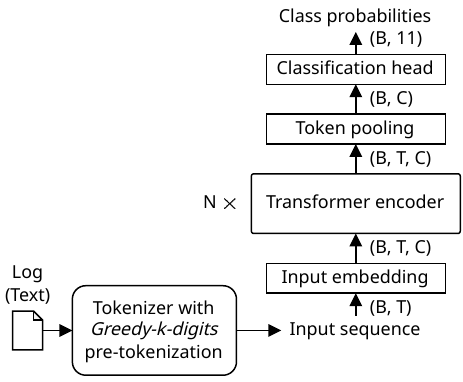}}
\caption{Model architecture}
\label{fig:transformer-architecture}
\end{figure}

Fig.~\ref{fig:transformer-architecture} shows the model architecture.
Each box visualizes a processing step in the pipeline; the arrow labels indicate the shape of the tensors between the processing steps.
Processing takes place in batches of size~$B$; other shape parameters are the sequence length~$T$ and the embedding dimension~$C$.
The processing steps are as follows:
\begin{itemize}
    \item \emph{Tokenizer with Greedy-k-digits pre-tokenization}:
The role of the tokenizer is to convert log texts of different content and length into a sequence of integers of length~$T$.
This conversion starts with pre-tokenization, which splits the input string into smaller substrings.
In natural language processing, pre-tokenization often splits on whitespace characters and punctuation marks, which we discovered through experimentation harms model performance.
Instead, we encountered that another simple pre-tokenization method works better for our use case, which we refer to as \emph{Greedy-k-digits}. 

\emph{Greedy-k-digits} splits strings whenever it encounters a new-line character or a digit sequence (which can optionally start with a minus-character) according to the following rules:
Splits are performed before the digit sequence, after the digit sequence, and after every $k$ digits within the sequence.
All digits are included in the output, but new-line characters are removed to reduce the output length.
The pre-tokenizer is called \emph{Greedy-k-digits} because it consumes as many digits as possible (up to $k$) before it introduces the split.

The intuition behind this tokenizer is that for many networking tools, two separate executions of the same tool cause output lines to differ only in a few words (if at all) or in an arbitrary amount of numbers.
Therefore---except for the numbers---the set of different lines that is encountered in the log files is small and complete lines (unless they contain any numbers) can be compressed into a single token.
When lines contain numbers, the lines must be split and numbers must be represented as tokens on their own.
Numbers should also be split after a certain number of digits.
A split after $k$~digits (with a small $k$) ensures that numbers are mapped to (combinations of) tokens that repeat in the data.
Without the split, numbers can be too unique to reappear multiple times and generalization becomes impossible.

After splitting, the sequence of tokens is truncated or padded with a special padding token to length~$T$.
Each of the $T$~tokens is mapped to an integer token identifier via a lookup table.
Because processing takes place in batches of size~$B$, this step produces a tensor of shape $(B, T)$.

    \item \emph{Input embedding}:
For each batch, an embedding is performed on each of the $T$~token identifiers, i.e. each token is converted to a $C$-dimensional vector.
Also, for each token, a second $C$-dimensional vector is created, which encodes the position of the token in the sequence.
The two vectors are summed up, resulting in the input embedding for the token.
The embeddings are learned during training via backpropagation.
The output of this step is a tensor of shape $(B, T, C)$.

    \item \emph{Transformer encoder}:
The main element of the architecture are the $N$ transformer encoder blocks, which follow the original transformer architecture \cite{vaswani2017} (including multi-head attention, skip connections, layer normalization, and dropout).
$N$ is a hyperparameter that has to be chosen manually or via automatic hyperparameter search.
The output tensor has shape $(B, T, C)$.

    \item \emph{Token pooling}:
For each batch, the sequence of $T$~token vectors (each of dimension~$C$) is summarized into a single vector of dimension~$C$. 
This reduction is performed using average pooling: the $i$-th entry of the output vector is the mean of the $i$-th entries of the $T$~token vectors.
The output of this token pooling step is a $(B, C)$ tensor.
    \item \emph{Classification head}:
Lastly, the tensor is passed into a linear layer with softmax activation.
This last layer consists of 11~units, each representing one of the classes.
Softmax activation ensures that the 11~values are valid probabilities.
The result of this final layer is a $(B, 11)$ tensor.
\end{itemize}

The training procedure and our hyperparameter choices are explained as part of our experiments in section~\ref{sec:evaluation}.

\section{Home Network Simulation and Data Generation}
\label{sec:simulation}

To train a classifier with supervised learning, we need a labeled dataset consisting of log files collected during problem situations and the corresponding problem class labels.
In principle, service providers who are in control of the residential gateways of their customers could obtain such a dataset by linking the residential gateway logs with the resolved trouble tickets of their technical support department.
For the purpose of this paper, we take a pragmatic approach and generate suitable data in a realistic simulator.

We implemented a simulation using Mininet-WiFi~\cite{fontes2015}, which is a fork of the original Mininet simulator extended by WiFi features.
Mininet-WiFi leverages Linux network namespaces to simulate multiple hosts on a single Linux machine.
Link properties like delay, packet loss, and jitter are simulated with \emph{tc}, a tool for manipulating traffic control settings.
As the simulation runs on an ordinary Linux system, it can use the same tools and libraries that are used on real routers and servers.
For instance, in our simulation, we utilize the widely-used dnsmasq DNS-server and the hostapd WiFi access point daemon. 
Mininet-WiFi supports the simulation of WiFi signal strengths via configurable propagation models.

\begin{figure}[b]
\centerline{\includegraphics[width=\linewidth]{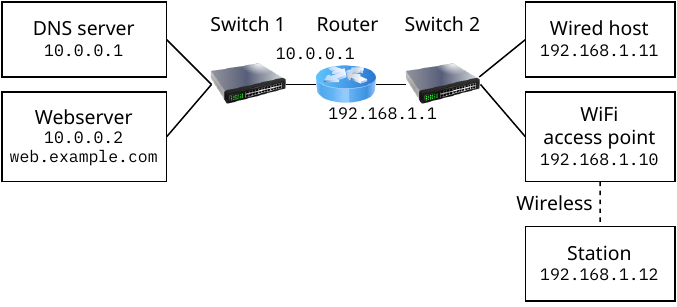}}
\caption{Simulation scenario}
\label{fig:simulation-scenario}
\end{figure}

Our simulation implements the topology of Fig.~\ref{fig:simulation-scenario}, which consists of the home network with a WiFi access point, a wireless client (a \emph{station} in IEEE~802.11 terminology), and a wired host.
This network is connected via a router to the (simulated) Internet, consisting of a DNS server and a webserver.
The webserver serves 14~images of different sizes via HTTP, each under their own URL.

Each simulation run is initialized with random variations to create heterogeneous behavior:
\begin{itemize}
    \item \emph{Random distance between WiFi access point and station}: The station is placed at different distances to the WiFi access point (between 0\,m and 10\,m), which affects the signal strength.
    \item \emph{Random link delay and jitter}:
Delay and jitter of the link between the wired host and Switch~2 are randomized (delay between 0\,ms and 100\,ms, jitter between 0\,ms and 20\,ms).
\item \emph{Random WiFi access point tx power}:
WiFi access point tx power is set randomly to values between 15\,dBm and 22\,dBm.
\end{itemize}

User activity is simulated by sending HTTP requests from the station to the webserver:
The station downloads between 1~and~5 random images from the webserver.
When the downloads are finished, the stations waits for a random interval between 100\,ms and 1000\,ms and requests another set of images randomly.
This leads to diverse traffic patterns in the logs.

The problem classes are simulated as follows:
\begin{itemize}
\item \emph{WiFi-related problems} are provoked by moving the station far away from the access point (\texttt{\footnotesize STATION\_FAR\_AWAY}) or by reducing the access point transmission power (\texttt{\footnotesize LOW\_AP\_TX\_POWER}).
\item \emph{Routing-related problems} are provoked by deleting the default gateway in the host's routing table (\texttt{\footnotesize NO\_DEFAULT\_ROUTE}) or by pointing it to a wrong IP address (\texttt{\footnotesize CORRUPT\_DEFAULT\_ROUTE}).
\item \emph{Link-related problems} are simulated by shutting down the ethernet interface on the host (\texttt{\footnotesize HOST\_INTERFACE\_DOWN}) or the router (\texttt{\footnotesize ROUTER\_INTERFACE\_DOWN}).
Problematic links with high packet loss (\texttt{\footnotesize HIGH\_PACKET\_LOSS}), high delay (\texttt{\footnotesize HIGH\_DELAY}), and high jitter (\texttt{\footnotesize HIGH\_JITTER}) are simulated using the built-in capabilities of Mininet-WiFi.
\item \emph{DNS-related problems} are simulated by making the DNS server unreachable by misconfiguration of the DNS server IP address in \texttt{/etc/resolv.d} (\texttt{\footnotesize DNS\_WRONG\_IP}).
\end{itemize}

The network monitoring application for data collection is implemented as a script that invokes ping, dig, ip, ethtool, and hostapd\_cli, concatenates the output of these tools, and writes it to a text file.
Between each script call, there is a 500\,ms pause.
We choose an intentionally short pause to generate many log files quickly.
Whenever a log file is produced, we store the currently simulated problem state as its label.
With this procedure, we obtain a labeled dataset that we can use to train and evaluate our model.

\section{Evaluation}
\label{sec:evaluation}

The goal of our evaluation is to demonstrate that the transformer model can reliably classify the 11~problem classes.
To quantify the classification performance, we 
measure accuracy, precision, and recall, which are common metrics for multi-class classification \cite{grandini2020}.

We evaluate the transformer model with \emph{Greedy-k-digits} pre-tokenization for $k=1,2,3,4$. 
We compare these models to a transformer with a whitespace pre-tokenizer that splits on whitespaces. 

Additionally, we compare our approach to a simple bag-of-words (BoW) model to investigate which problem classes can be identified with BoW and which classes require the more complex transformer model.
We choose BoW as a baseline because it can work on text directly, in contrast to many other algorithms that require parsing and feature engineering.
In the BoW model, the input tokens are embedded into a multi-hot vector $\mathbf{x}$, where $x_i = 1$ if the $i$-th token of the vocabulary is present in the input tokens and $x_i = 0$ otherwise. 
The multi-hot vector is passed to a linear layer of 11~units with softmax activation.
We report metrics for BoW with a whitespace pre-tokenizer and with \emph{Greedy-k-digits} for $k=3$.

In summary, we study the following seven test cases:
\begin{itemize}
    \item Our \emph{proposed method} as described in section~\ref{sec:approach}:
    \begin{itemize}
        \item[1.] Transformer with \emph{Greedy-k-digits} and $k=1$
        \item[2.] Transformer with \emph{Greedy-k-digits} and $k=2$
        \item[3.] Transformer with \emph{Greedy-k-digits} and $k=3$
        \item[4.] Transformer with \emph{Greedy-k-digits} and $k=4$
    \end{itemize}
    \item \emph{Baselines} for comparison:
    \begin{itemize}
        \item[5.] Transformer with whitespace pre-tokenizer
        \item[6.] BoW model with \emph{Greedy-k-digits} and $k=3$
        \item[7.] BoW model with whitespace pre-tokenizer
    \end{itemize}
\end{itemize}

\subsection{Experiment Setup}

For model training, we generate a dataset with the simulation described in section~\ref{sec:simulation}.
The dataset consists of $356{,}061$ samples with an approximately equal amount of samples per class, which we split into a train, validation, and test set:
\begin{itemize}
    \item \emph{Train set}: $246{,}241$ samples
    \item \emph{Validation set}: $61{,}860$ samples
    \item \emph{Test set}: $47{,}960$ samples
\end{itemize}
The train and validation set are used for model training and hyperparameter tuning, the test set for this final evaluation.

The transformer model can be trained with different hyperparameter values, which affect the model size, training time, and model accuracy.
During model development, we found the best hyperparameter values through experimental exploration on the validation set.
We tried smaller models and increased the model capacity by adding more layers or increasing layer widths until it stopped improving our results.
The hyperparameter values that we use for evaluation on the test set are listed in Table~\ref{tab:hyperparameters}.

\begin{table}[ht]
\centering
\caption{Hyperparameters}
\label{tab:hyperparameters}
\begin{tabular}{llr}
\toprule
Model & Hyperparameter & Value \\ \midrule
Transformer & Optimizer & AdamW \\
& Learning rate &  \num{1e-5}    \\
& Batch size $B$ & $64$  \\
& Sequence length $T$ & $512$ / $1024$ \\
& Embedding dim. $C$ & $64$ \\
& Transformer blocks $N$ & $3$ \\
& Attention heads &  $2$ \\
& Transformer intermediate dim. & $128$ \\
& Layer norm. epsilon & \num{1e-5} \\
& Dropout rate & $10\%$ \\
BoW model & Optimizer & AdamW \\
& Learning rate  &  \num{1e-4}     \\
& Batch size   &   $64$    \\ \bottomrule
\end{tabular}
\end{table}

We use different sequence lengths~$T$ for the two pre-tokenizers:
The whitespace pre-tokenizer introduces more splits than the \emph{Greedy-k-digits} pre-tokenizer and requires a larger sequence length~$T$ to fit all tokens.
We use a sequence length~$T=1024$ for the whitespace pre-tokenizer and $T=512$ for the \emph{Greedy-k-digits} pre-tokenizer.

All computations are performed on a \emph{g4dn.xlarge} instance on AWS with an NVIDIA T4 GPU, 4 vCPUs, and 16GiB of memory.
All of the models can be trained within a few hours.

\subsection{Experiment Results}
\label{subsec:evaluation}

Table~\ref{tab:evaluation} shows for each of the seven test cases the accuracy as well as the precision and recall for each problem class.
If one model uniquely achieves the best precision or recall value for a problem class, we format these values \textbf{\color{JungleGreen}green}.
Similarly, if one model uniquely has the worst precision or recall value for a problem class, we format these values \textbf{\color{RedOrange} red}.
If multiple models share the best or worst value, we format them in black standard font.

\begin{table*}
\centering
\caption{Experiment results}
\label{tab:evaluation}
\begin{tabular}{@{}lcccccccccccccc@{}}
\toprule
& \multicolumn{2}{c}{\begin{tabular}[c]{@{}c@{}}Transformer\\$k=1$\end{tabular}}
& \multicolumn{2}{c}{\begin{tabular}[c]{@{}c@{}}Transformer\\$k=2$\end{tabular}}
& \multicolumn{2}{c}{\begin{tabular}[c]{@{}c@{}}Transformer\\$k=3$\end{tabular}}
& \multicolumn{2}{c}{\begin{tabular}[c]{@{}c@{}}Transformer\\$k=4$\end{tabular}}
& \multicolumn{2}{c}{\begin{tabular}[c]{@{}c@{}}Transformer\\Whitespace\end{tabular}}
& \multicolumn{2}{c}{\begin{tabular}[c]{@{}c@{}}BoW model\\$k=3$\end{tabular}}
& \multicolumn{2}{c}{\begin{tabular}[c]{@{}c@{}}BoW model\\Whitespace\end{tabular}}\\
\midrule
\textbf{Precision (P) / Recall (R)} & P & R
& P & R
& P & R
& P & R
& P & R
& P & R
& P & R \\ 
\cmidrule[0.5pt](lr){2-3}
\cmidrule[0.5pt](lr){4-5}
\cmidrule[0.5pt](lr){6-7}
\cmidrule[0.5pt](lr){8-9}
\cmidrule[0.5pt](lr){10-11}
\cmidrule[0.5pt](lr){12-13}
\cmidrule[0.5pt](lr){14-15}
\texttt{\footnotesize CORRUPT\_DEFAULT\_ROUTE}            &1.00	 & 1.00	 & 1.00	 & 1.00	 & 1.00	 & 1.00	 & 1.00	 & 1.00	 & 1.00	 & 1.00	 & 1.00	 & 1.00	 & 1.00	 & 1.00 \\
\texttt{\footnotesize DNS\_WRONG\_IP}       &1.00	 & 1.00	 & 1.00	 & 1.00	 & 1.00	 & 1.00	 & 1.00	 & 1.00	 & 1.00	 & 1.00	 & 1.00	 & 1.00	 & \textbf{\color{RedOrange} 0.96}	 & \textbf{\color{RedOrange} 0.99} \\
\texttt{\footnotesize HIGH\_DELAY}  &1.00	 & 1.00	 & 1.00	 & 1.00	 & 1.00	 & 1.00	 & 1.00	 & 1.00	 & 1.00	 & 1.00	 & 1.00	 & 1.00	 & \textbf{\color{RedOrange} 0.99}	 & 1.00 \\
\texttt{\footnotesize HIGH\_JITTER}           &0.81	 & 0.88	 & 0.93	 & 0.99	 & $\mathbf{\color{JungleGreen}0.99}$	 & 0.99	 & 0.96	 & 0.99	 & 0.68	 & 0.96	 & 0.84	 & 0.79	 & \textbf{\color{RedOrange} 0.67}	 & \textbf{\color{RedOrange} 0.73} \\
\texttt{\footnotesize HIGH\_LOSS}       &1.00	 & 1.00	 & 1.00	 & 0.99	 & 1.00	 & 1.00	 & 1.00	 & 1.00	 & \textbf{\color{RedOrange} 0.92}	 & 1.00	 & 0.98	 & 0.99	 & 0.94	 & 1.00 \\
\texttt{\footnotesize HOST\_INTERFACE\_DOWN}       &1.00	 & 1.00	 & 1.00	 & 1.00	 & 1.00	 & 1.00	 & 1.00	 & 1.00	 & 1.00	 & 1.00	 & 1.00	 & 1.00	 & 1.00	 & 1.00 \\
\texttt{\footnotesize LOW\_AP\_TX\_POWER}    &0.61	 & \textbf{\color{RedOrange} 0.42}	 & 0.69	 & 0.67	 & 0.69	 & $\mathbf{\color{JungleGreen}0.71}$	 & 0.68	 & 0.68	 & 0.50	 & 0.51	 & 0.63	 & 0.66	 & \textbf{\color{RedOrange} 0.39}	 & 0.43 \\
\texttt{\footnotesize NO\_DEFAULT\_ROUTE}  &1.00	 & 1.00	 & 1.00	 & 1.00	 & 1.00	 & 1.00	 & 1.00	 & 1.00	 & 1.00	 & 1.00	 & 1.00	 & 1.00	 & 1.00	 & 1.00 \\
\texttt{\footnotesize NORMAL\_STATE}       &0.77	 & 0.67	 & 0.98	 & 0.68	 & $\mathbf{\color{JungleGreen}0.99}$	 & 0.72	 & 0.90	 & 0.74	 & 0.60	 & 0.49	 & 0.78	 & 0.66	 & \textbf{\color{RedOrange}0.40}	 & \textbf{\color{RedOrange} 0.45} \\
\texttt{\footnotesize ROUTER\_INTERFACE\_DOWN}              &1.00	 & 1.00	 & 1.00	 & 1.00	 & 1.00	 & 1.00	 & 1.00	 & 1.00	 & 1.00	 & 1.00	 & 1.00	 & 1.00	 & \textbf{\color{RedOrange} 0.99}	 & \textbf{\color{RedOrange} 0.96} \\
\texttt{\footnotesize STATION\_FAR\_AWAY}             &0.72	 & $\mathbf{\color{JungleGreen}0.98}$	 & 0.77	 & 0.97	 & 0.77	 & 0.96	 & 0.80	 & 0.91	 & $\mathbf{\color{JungleGreen}0.85}$	 & 0.56	 & 0.79	 & 0.91	 & \textbf{\color{RedOrange} 0.64}	 & \textbf{\color{RedOrange} 0.40} \\

\midrule
\textbf{Accuracy}
& \multicolumn{2}{c}{0.90}
& \multicolumn{2}{c}{0.94}
& \multicolumn{2}{c}{0.94}
& \multicolumn{2}{c}{0.94}
& \multicolumn{2}{c}{0.87}
& \multicolumn{2}{c}{0.91}
& \multicolumn{2}{c}{0.81}\\

\bottomrule
\end{tabular}
\end{table*}

Based on Table~\ref{tab:evaluation}, we make the following observations:
\begin{itemize}
    \item \emph{Some problem classes can be correctly identified with $100\%$ accuracy by all models.}
    These problem classes can be identified by the presence and absence of certain keywords and are hence easily identifiable for all models.
    For instance, all models can reliably identify if the host interface is down or no default route is configured.
    If only this subset of classes is of interest, a simple BoW model is preferable to the more complex transformer as the BoW can be trained faster.
    \item \emph{Greedy-k-digits pre-tokenizer beats whitespace pre-tokenizer in all seven test cases.}
    The lowest accuracy obtained with \emph{Greedy-k-digits} is $0.90$, which is higher than the highest accuracy obtained with the whitespace pre-tokenizer ($0.87$).
    This supports our claim that whitespace pre-tokenization is not ideal for splitting tool output and special treatment of numbers improves model accuracy from at least $0.87$ to $0.90$.
    \item \emph{The best models are the transformers with Greedy-k-digits for} $k=2,3,4$.
    These three transformer models achieve an accuracy of $0.94$.
    It is noteworthy that the transformer with $k=3$ achieves as only model $0.99$ precision for \texttt{\footnotesize HIGH\_JITTER}, $0.71$ recall for \texttt{\footnotesize LOW\_AP\_TX\_POWER}, and $0.99$ precision for \texttt{\footnotesize NORMAL\_STATE}.
    For $k=1$, only $0.90$ accuracy is achieved and in particular the recall value of $0.42$ for \texttt{\footnotesize LOW\_AP\_TX\_POWER} is the lowest observed recall value for that class.
    \item \emph{The worst model is BoW with whitespace pre-tokenizer.}
    This model only achieves $0.81$ accuracy and for many classes has the worst precision and recall.
    In almost all cases this model can be improved by replacing the whitespace pre-tokenizer with the \emph{Greedy-k-digits} pre-tokenizer.
\end{itemize}

\begin{figure*}
     \centering
     \begin{subfigure}[b]{0.49\textwidth}
         \centering
         \includegraphics[trim={0.25cm 0.25cm 0 0.3cm},clip,width=\textwidth]{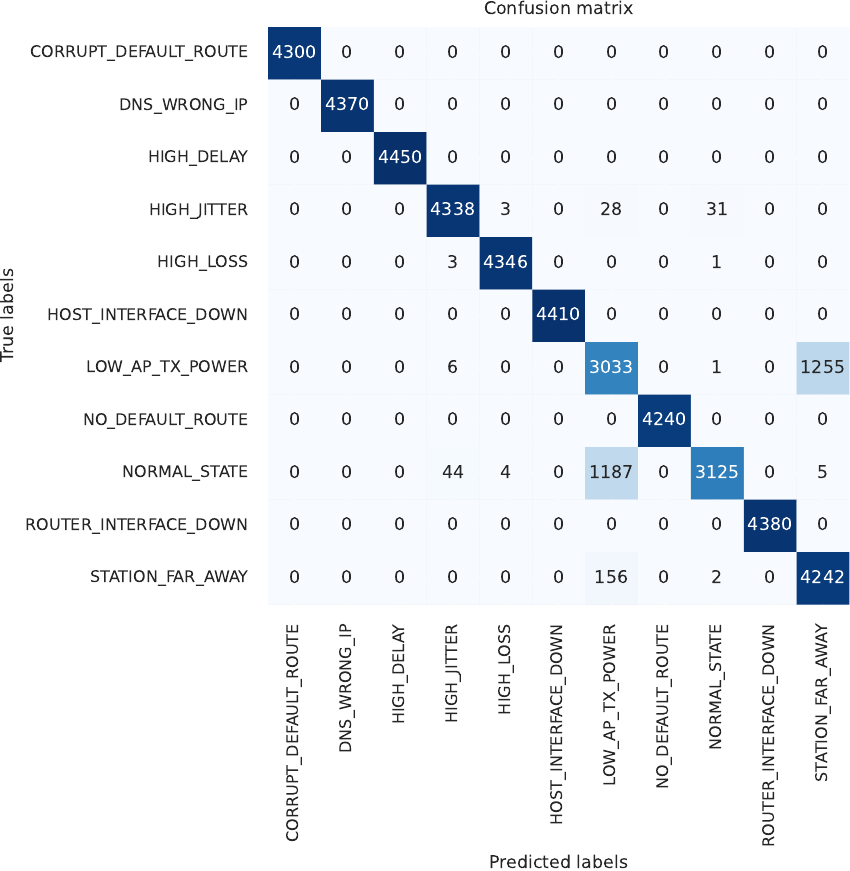}
         \caption{Transformer with \emph{Greedy-k-digits} for $k=3$}
         \label{fig:confusion-matrix-transformer}
     \end{subfigure}
     \begin{subfigure}[b]{0.49\textwidth}
         \centering
         \includegraphics[trim={0.25cm 0.25cm 0 0.3cm},clip,width=\textwidth]{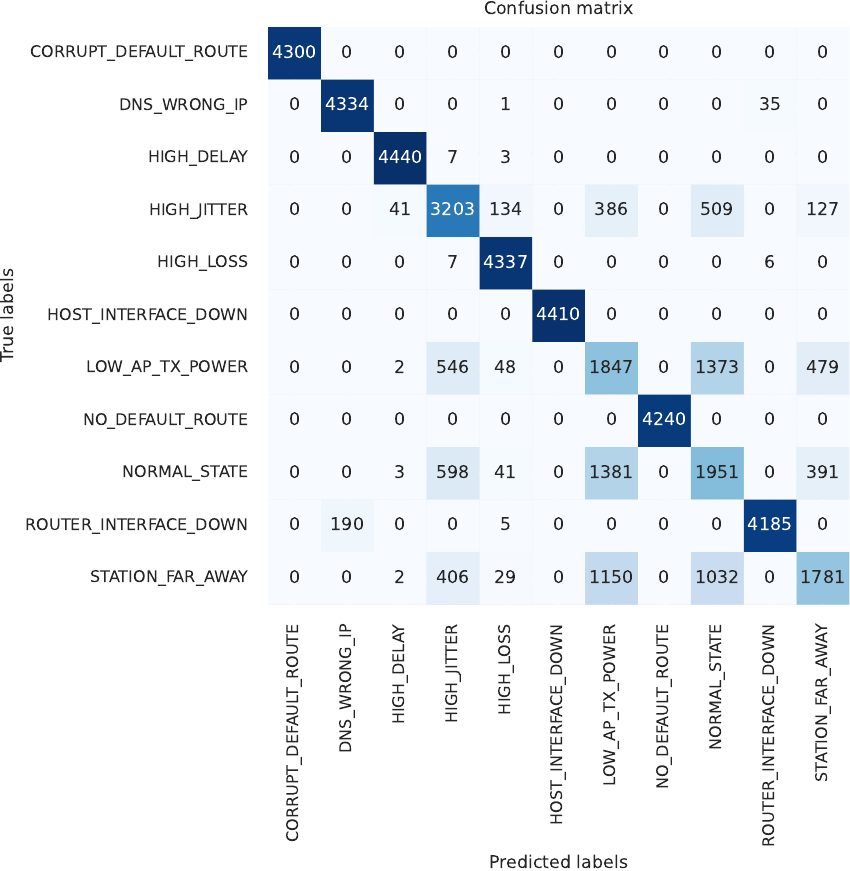}
         \caption{BoW model with whitespace tokenizer}
         \label{fig:confusion-matrix-bow}
     \end{subfigure}
     \hfill
      \caption{Confusion matrix\\{\normalfont (Predicted labels on the x-axis / true labels on the y-axis)}}
        \label{fig:confusion-matrix}
\end{figure*}

To analyze the classification mistakes more deeply, we compare the confusion matrices of the transformer with \emph{Greedy-k-digits} for $k=3$ and the BoW model with whitespace pre-tokenizer, which are shown in Fig.~\ref{fig:confusion-matrix}.
The predicted labels are plotted on the x-axis and the true labels on the y-axis.
\begin{itemize}
    \item \emph{Transformer confusion matrix} (Fig.~\ref{fig:confusion-matrix-transformer}):
    The main mistakes made by the transformer are related to the \texttt{\footnotesize LOW\_AP\_TX\_POWER} class.
    For log files of this problem class, the transformer sometimes wrongly predicts \texttt{\footnotesize STATION\_FAR\_AWAY}.
    This can be explained by the fact that both problem classes reduce the signal strength of the station and there seems to be no clear pattern in the logs for the transformer to distinguish between the two classes.
    Also, the transformer sometimes misclassifies \texttt{\footnotesize NORMAL\_STATE} logs as \texttt{\footnotesize LOW\_AP\_TX\_POWER}.
    It seems likely that in some cases the low AP tx power affects the signal strength only slightly and the transformer cannot learn a clear distinction between the two classes.
    \item \emph{BoW confusion matrix} (Fig.~\ref{fig:confusion-matrix-bow}):
    The BoW model commits many errors for \texttt{\footnotesize HIGH\_JITTER}, \texttt{\footnotesize LOW\_AP\_TX\_POWER}, \texttt{\footnotesize NORMAL\_STATE}, and \texttt{\footnotesize STATION\_FAR\_AWAY}.
    For these four classes, the BoW model produces many false positives and false negatives, which strongly reduces the model's utility for these classes.
    Interestingly, the BoW model performs poorly for \texttt{\footnotesize HIGH\_JITTER}, which the transformer model can identify almost flawlessly.
    For detecting high jitter, the model must identify the delay measurement values in the log and recognize that they have high variance.
    This capability of the transformer is a clear advantage over the BoW model.
    Additionally, the BoW model occasionally misclassifies \texttt{\footnotesize ROUTER\_INTERFACE\_DOWN} as \texttt{\footnotesize DNS\_WRONG\_IP}.
    Here the BoW model is likely misled by the fact that both classes report failed DNS queries in the log.
    In contrast, the transformer can handle this difficulty without any problems.
\end{itemize}

In summary, our transformer model with \emph{Greedy-k-digits} and $k=3$ achieves the best results in our experiments and shows the benefits of transformers for the more difficult problem classes compared to the BoW model.
The \emph{Greedy-k-digits} pre-tokenizer consistently outperforms the whitespace pre-tokenizer in our setting.

\section{Conclusion}
\label{sec:conclusion}

We have presented a transformer-based deep learning model for classification of common home network problems.
The model can identify ten problems accurately based on the raw textual output of networking tools without the need for cumbersome parsing or feature engineering.
While simple problem classes can be classified with simpler bag-of-words models, our transformer model is applicable to a wider range of problems and can also identify the more difficult classes reliably.
The \emph{Greedy-k-digits} pre-tokenizer that we proposed in the paper supports the models in achieving high accuracy, precision, and recall and outperforms whitespace pre-tokenization in all of the evaluated test cases.

Model training and evaluation was performed on simulation data.
While additional data from field deployments would be valuable to further evaluate our model, the simulation data should be sufficiently realistic to showcase the large potential of our approach.
For model deployment, the model should be trained on logs and labels acquired from field networks and resolved trouble tickets.

Future work could study how transformer-based network problem classification can be transferred to other networking areas, e.g., optical access networks or mobile networks.
Also, our solution approach is not limited to the demonstrated 10~problem classes and can be expanded to additional classes.
A variant of our transformer model might be a useful building block for future machine-learning-based troubleshooting engines and self-healing networks.

Network problem classification models like the one presented in this paper are of high practical relevance:
Once a problem is accurately classified by a model, the problem can be explained to the subscriber or in some cases even resolved automatically.

\section*{Acknowledgments}
This work has been partially funded by the German Federal Ministry of Education and Research (BMBF) in the FRONT-RUNNER project (Grant 16KISR005K).

\bibliographystyle{ACM-Reference-Format}
\bibliography{bib}

\end{document}